\documentclass[aps,pre,twocolumn,showpacs,preprintnumbers,amsmath,amssymb,floatfix]{revtex4}
\usepackage{epsfig}
\input epsf
\epsfclipon
\usepackage{epstopdf}

\begin{document}
\title{Exponential random graph models for networks with community structure}
%\title{Blockmodels: an exponential random graph approach}
%\title{Globalization puzzle in the gravity law of trade}
\author{Piotr Fronczak, Agata Fronczak, and Maksymilian Bujok}
\affiliation{Faculty of Physics, Warsaw University of Technology,
Koszykowa 75, PL-00-662 Warsaw, Poland}
\date{\today}

\begin{abstract}
Although the community structure organization is one of the most important
characteristics of real-world networks, the traditional network models fail
to reproduce the feature. Therefore, the models are useless as benchmark
graphs for testing community detection algorithms. They are also inadequate
to predict various properties of real networks. With this paper we intend to
fill the gap. We develop an exponential random graph approach to networks
with community structure. To this end we mainly built upon the idea of
blockmodels. We consider both, the classical blockmodel and its
degree-corrected counterpart, and study many of their properties
analytically. We show that in the degree-corrected blockmodel, node degrees
display an interesting scaling property, which is reminiscent of what is
observed in real-world fractal networks. The scaling feature comes as a
surprise, especially that in this study, contrary to what is suggested in the
literature, the scaling property is not attributed to any specific network
construction procedure. It is an intrinsic feature of the degree-corrected
blockmodel. A short description of Monte Carlo simulations of the models is
also given in the hope of being useful to others working in the field.
\end{abstract}  \maketitle

\section{Introduction}

Of particular interest in recent years has been the community structure in
networks \cite{PhysRep2010Fortunato, NaturePhys2012Newman}. As a community
(module or block), one understands a group of nodes that is densely connected
internally but sparsely connected externally. To name a few, communities may
be groups of related individuals in social networks, sets of web-pages on the
same topic, biochemical pathways in metabolic networks, and groups of
countries in the world trade network that signed regional trade agreements.
The above examples show that the group membership is related to the function
of a node in the network. For this reason, aside from the small-world effect
and scale-free degree distributions, the community structure is considered
one of the most important topological properties of complex networks, yet
this structure is not fully understood and not well captured by network
models.

One of the proposed models for networks with community structure, with a long
tradition of study in the social sciences and computer science
\cite{SocNet1983Holland, SocNet1992Faust, SocNet1992Anderson,
JClass1997Snijders, JMach2008Airoldi, JMach2009Goldenberg}, is the so-called
blockmodel. In its classical version \cite{SocNet1983Holland}, each of $N$
vertices is assigned to one of $K$ blocks (communities) of equal size, and
undirected edges are independently drawn between pairs of nodes with
probabilities that are a function only of the group membership of the nodes.
The well-defined community structure of this model caused it to be the most
regularly used benchmark for testing community detection algorithms
\cite{PNAS2002Girvan, PRE2004Newman}.

However, apart from the built-in blocks (communities), other properties of
this model, especially the Poisson-like degree distribution, indicate its
shortcomings as the correct model for real networks. To our knowledge, in
respect to community detection algorithms, the concern was first raised by
Lancichinetti, Fortunato, and Radicchi \cite{PRE2009Lancichinetti}. The
authors argued that real networks are characterized by heterogeneity in the
distributions of node degrees and of community sizes, which is not the case
in the classical blockmodel. With this motivation Lancichinetti et al. have
proposed an efficient numerical construction procedure for benchmark graphs
that account for the desired network properties.

More recently, a similar perspective was also raised in other papers, see
e.g. \cite{PRE2011Karrer, PRE2011Shen, PRE2012Peixoto, PRE2012Seshadhri,
PhysA2012Gregory}. In particular, Karrer and Newman in
Ref.~\cite{PRE2011Karrer} showed that due to limitations of the traditional
blockmodel, its fitting to empirical network data, which is a way of
discovering community structure, may be misleading. The authors suggested,
how to generalize the classical blockmodel to incorporate arbitrary degree
distributions. They showed that the degree-corrected counterpart dramatically
outperforms the traditional blockmodel as a tool for detecting community
structure.

In the following, we present an exponential random graph formulation
\cite{PRE2004Park, PRE2006Fronczak, inbookNewman, Rev2013Shalizi,
Essay2012Fronczak} for both the traditional and the degree-corrected
blockmodels. The aim behind is to create a general model for networks with
community structure, which has a formal mathematical foundation and is easy
to implement in numerical simulations. Models of networks are very important
in the study of network processes and algorithms. As already indicated,
community detection algorithms can be evaluated more effectively on synthetic
networks with a well-defined community structure (benchmarks) than on real
networks \cite{PRE2009Lancichinetti}, because one can easily vary the model
parameters and compare the recovered community structure with the predefined
one. Network models are also ubiquitous in studies of different processes
that takes place on networks \cite{bookBarrat}, such as the spread of a
disease over a social network, or the flow of traffic on communication
networks. Therefore, it is quite surprising that most of the current network
models disregard the issue of community structure. This paper intends to fill
the gap.

%tutaj trzeba omówiæ co my dok³adnie pokazaliœmy. Jakie relacje znaleŸliœmy dla sieci degree-corrected fraktalnoœæ sieci. itp...

The outline of the paper is as follows. In Sec.~\ref{ClassicalB}, the
classical blockmodel is considered based on exponential random graph
approach. In Sec.~\ref{DegreeB}, its degree-corrected version is studied and
various network properties are calculated. In Sec.~\ref{MonteCarloB}, a short
training in Monte Carlo simulations, to which the considered models lend
themselves admirably, is given. The paper is concluded in
Sec.~\ref{Conclude}.

\section{Classical blockmodel}\label{ClassicalB}

Exponential random graphs (ERGs) are ensemble models, which are defined not
to be a single network but a collection, $\{G\}$, of possible networks. A
graph $G$ in the ensemble is assigned the probability
\begin{equation}\label{PG0}
P(G)=\frac{e^{H(G)}}{Z},
\end{equation}
where
\begin{equation}\label{Z0}
Z=\sum_{\{G\}}e^{H(G)}
\end{equation}
is the normalization constant (partition function) and
\begin{equation}\label{H0}
H(G)=\sum_{r}\theta_rm_r(G)
\end{equation}
is called the graph Hamiltonian, with $\{m_r(G)\}$ being a collection of
graph observables (which reflect desired properties of the model), and
$\{\theta_r\}$ standing for model parameters (which are coupled to
observables).

In this paper, when we will talk about possible realizations, $\{G\}$, of
networks with community structure we will always refer to simple graphs with
$N$ labeled nodes and independent edges. To state precisely, a simple graph,
$G$, has at most one link between any pair of nodes and it does not contain
self-loops connecting nodes to themselves. It means that entries of its
adjacency matrix, $A_{ij}(G)\equiv A_{ij}$ and $A_{ij}\in\{0,1\}$, are
symmetric, $A_{ij}=A_{ji}$, and they are equal to zero on the diagonal,
$A_{ii}=0$. Therefore, by assuming statistical independence of the links, the
probability of a graph, $G$, can be written as
\begin{equation}\label{PG1}
P(G)=\prod_{i<j}p_{ij}^{A_{ij}}(1-p_{ij})^{1-A_{ij}},
\end{equation}
where $p_{ij}$ is the probability that the vertices $i$ and $j$ are
connected.

It is remarkable, that Eq.~(\ref{PG1}) can be rewritten in the canonical form
of Eq.~(\ref{PG0}),
\begin{equation}\label{PG2}
P(G)=\frac{\exp\left[\sum_{i<j}\ln\left(\frac{p_{ij}}{1-p_{ij}}\right)
A_{ij}\right]}{\prod_{i<j}(1-p_{ij})^{-1}},
\end{equation}
with the Hamiltonian defined as
\begin{equation}\label{HG1}
H(G)=\sum_{i<j}\theta_{ij}A_{ij},
\end{equation}
where
\begin{equation}\label{thetaijG1}
\theta_{ij}=\ln\left(\frac{p_{ij}}{1-p_{ij}}\right),
\end{equation}
is a separate parameter (the so-called Lagrange multiplier) coupling to an
edge between $i$ and $j$, and the partition function is given by
\begin{equation}\label{ZG1}
Z(\{\theta_{ij}\})=\prod_{i<j}(1-p_{ij})^{-1}=\prod_{i<j}(1+e^{\theta_{ij}}).
\end{equation}
Furthermore, one can show that the expected value of each matrix entry,
$\langle A_{ij}\rangle$, can be calculated by differentiating the logarithm
of the partition function, $\ln Z(\{\theta_{ij}\})$, with respect to
$\theta_{ij}$,
\begin{equation}\label{Aij1}
\langle A_{ij}\rangle=\frac{\partial\ln Z}{\partial\theta_{ij}}=\frac{e^{\theta_{ij}}}{1+e^{\theta_{ij}}}=p_{ij}.
\end{equation}

In the classical blockmodel, one assumes that the number of groups, $K$, and
the number of vertices in each group, $N_r$, are known, and
\begin{equation}\label{BNr}
\sum_{r=1}^KN_r=N,
\end{equation}
where $N$ is the total number of vertices in the network \footnote{Let us
note that in the classical blockmodel all communities are of the same size,
see Refs.~\cite{SocNet1983Holland, SocNet1992Faust, SocNet1992Anderson,
JClass1997Snijders}, differently from what is assumed here. A variant of the
the standard blockmodel with communities of different size was first
introduced by Donan et al. in Ref.~\cite{JStatMech2005Danon}.}. One also
assumes that the probability that the vertices $i$ and $j$ are connected is a
function only of their group membership,
\begin{equation}\label{Bpij}
p_{ij}\equiv q_{g_ig_j},
\end{equation}
where $g_i$ and $g_j$ represent the groups to which the vertices belong. With
this in mind, the parameter $\theta_{ij}$, Eq.~(\ref{thetaijG1}), can be
written as
\begin{equation}\label{Bthetaij}
\theta_{ij}=\ln\left(\frac{q_{g_ig_j}}{1-q_{g_ig_j}}\right)\equiv \omega_{g_ig_j},
\end{equation}
and the Hamiltonian of the classical blockmodel gets the informative form
\begin{eqnarray}\label{BH1}
H(G)\!&=&\!\sum_{i<j}\omega_{g_ig_j}A_{ij}\!=\!\sum_{r\leq s}\omega_{rs}\!\sum_{i<j}A_{ij}\delta_{g_ir}
\delta_{g_js}\\\label{BH3}\!&=&\!\sum_{r\leq s}\omega_{rs}E_{rs}(G),
\end{eqnarray}
where $E_{rs}(G)$ is the number of edges between the groups~$r$ and $s$,
$E_{rr}(G)$ represents edges within the same group $r$, and $\delta_{rs}$ is
the Kronecker delta.

The Hamiltonian obtained, Eq.~(\ref{BH3}), shows that the classical
blockmodel is equivalent to the ensemble of networks with the specified
average numbers of edges within and between the predefined blocks, i.e.
$\langle E_{rr}\rangle$ and $\langle E_{rs}\rangle$, respectively. The
partition function, Eq.~(\ref{ZG1}), corresponding to this ensemble can,
after an amount of algebra, be written in the convenient form
\begin{eqnarray}\label{BZ1}
Z(\{\omega_{rs}\})\!\!&=&\!\!\prod_{r<s}(1\!-\!q_{r\!s})^{-N_{\!r}\!N_{\!s}}
\prod_{r}(1\!-\!q_{r\!r})^{-{N_{\!r}\choose 2}}
\\\label{BZ2}\!\!&=&\!\!\prod_{r<s}(1\!+\!e^{\omega_{r\!s}})^{N_{\!r}\!N_{\!s}}\prod_{r}(1\!+\!e^{\omega_{r\!r}})^{{N_{\!r}\choose 2}},
\end{eqnarray}
that allows to calculate the mentioned averages:
\begin{eqnarray}\label{BErr}
\langle E_{rr}\rangle\!&=\!&\frac{\partial\ln Z}{\partial \omega_{rr}}=
{\!N_{\!r}\choose 2\!}\frac{e^{\omega_{\!rr}}}{1+e^{\omega_{\!rr}}}={\!N_{\!r}\choose 2\!}q_{rr},\\ \label{BErs}
\langle E_{rs}\rangle\!&=&\!\frac{\partial\ln Z}{\partial \omega_{rs}}=
N_{\!r}N_{\!s}\frac{e^{\omega_{\!rs}}}{1+e^{\omega_{\!rs}}}=N_{\!r}N_{\!s}q_{rs}.
\end{eqnarray}

Eqs.~(\ref{BErr}) and (\ref{BErs}) can be used to calculate internal and
external degrees of a node $i$, i.e. $\langle k_i^{int}\rangle$ and $\langle
k_i^{ext}\rangle$, which are the numbers of edges that connect the node to
other nodes in the same block and nodes from different blocks, respectively.
In order to do it, let us assume that the partial node's degree, $k_{i,r}^s$,
is the number of edges that are incident to $i$, where $g_i=r$, and such that
their second endpoint belongs to the block $s$, i.e.
\begin{equation}\label{kis}
k_{i,r}^s=\delta_{g_ir}\sum_{j}A_{ij}\delta_{g_js}.
\end{equation}
Then, the internal and external degrees can be written as
\begin{eqnarray}\label{kiint}
\langle k_i^{int}\rangle&=&\langle k_{i,r}^{r}\rangle,\\\label{kiext}
\langle k_i^{ext}\rangle&=&\sum_{s,s\neq r}\langle k_{i,r}^{s}\rangle,
\end{eqnarray}
and the average node's connectivity through the whole network is
\begin{equation}\label{ki}
\langle k_i\rangle=\left\langle\sum_sk_{i,r}^s\right\rangle=
\langle k_i^{int}\rangle+\langle k_i^{ext}\rangle.
\end{equation}
In the classical blockmodel, Eqs.~(\ref{kiint}) and~(\ref{kiext}) simplify
to:
\begin{equation}\label{Bkiint}
\langle k_i^{int}\rangle=\delta_{g_ir}\frac{2\langle E_{rr}\rangle}{N_{r}},
=\delta_{g_ir}(N_r-1)q_{rr},
\end{equation}
and
\begin{equation}\label{Bkiext}
\langle k_i^{ext}\rangle=\delta_{g_ir}\sum_{s,s\neq r}\frac{\langle E_{rs}\rangle}{N_{r}}
=\delta_{g_ir}\sum_{s,s\neq r}N_sq_{rs}.
\end{equation}

In Sec.~\ref{MonteCarloB}, the results from this section are used to
introduce and discuss the method of Monte Carlo simulations for networks with
community structure.

\section{Degree-corrected blockmodel}\label{DegreeB}

In this section, we develop an exponential random graph approach to networks
with community structure that accounts for the heterogeneity of both degree
and community size. We assume that the probability, $p_{ij}$, that vertices
$i$ and $j$ are connected is not only a function of their group membership
(as it was the case of the classical blockmodel), but also depends on the
vertices themselves. In doing so, we examine the Hamiltonian,
Eq.~(\ref{HG1}), with the following set of the edge parameters,
\begin{equation}\label{Dthetaij1}
\theta_{ij}=v_i+v_j+\omega_{g_ig_j},
\end{equation}
where the parameters $\{v_i\}$ are thought to control expected node degrees
and $\omega_{rs}$ controls edges between groups $r\!=\!g_i$ and $s\!=\!g_j$
to which the nodes $i$ and $j$ belong.

By inserting Eq.~(\ref{Dthetaij1}) into Eq.~(\ref{HG1}), the Hamiltonian of
the considered network ensemble can, after an amount of algebra, be written
in two alternative forms: the first consisting of a sum over the node degrees
and over the edges within and between the predefined blocks, i.e.
\begin{eqnarray}\label{DH1a}
H(G)\!&=&\!\sum_{i<j}\left(v_i A_{ij}\!+\!v_j A_{ji}\right)\!+\!\sum_{i<j}\omega_{g_ig_j}A_{ij}\\\label{DH1b}
\!&=&\!\sum_{i,j}v_i A_{ij}\!+\!\sum_{r\leq s}\omega_{rs}\!\sum_{i<j}\delta_{g_ir}\delta_{g_js}A_{ij}\\\label{DH1c}
\!&=&\!\sum_{i}v_i k_i(G)\!+\!\sum_{r\leq s}\omega_{rs}E_{rs}(G),
\end{eqnarray}
and the second being the sum over the partial node's degrees, i.e.
\begin{eqnarray}\label{DH2a}
H(G)\!&=&\!\sum_{i,j}\left(v_i\!+\!\frac{\omega_{g_ig_j}}{2}\right)A_{ij}\\\label{DH2b}
\!&=&\!\sum_{r,s}\!\sum_i\left(v_i\!+\!\frac{\omega_{rs}}{2}\right)k_{i,r}^s(G)
\\\label{DH2c}\!&=&\!\sum_{r,s}\!\sum_iv_{i,r}^{s}k_{i,r}^s(G)\\\label{DH2d}
\!&=&\!\sum_{r}\!\sum_i\!v_{i,r}^{r}k_{i,r}^r(G)\!+\!\sum_{r\neq s}\!\sum_i\!v_{i,r}^{s}k_{i,r}^s(G),
\end{eqnarray}
where the new set of parameters,
\begin{equation}\label{varthetai}
v_{i,r}^{s}=\left(v_i\!+\!\frac{\omega_{rs}}{2}\right),
\end{equation}
is introduced, such that the edge parameters, $\theta_{ij}$,
Eq.~(\ref{Dthetaij1}), can be written as
\begin{equation}\label{Dthetaij2}
\theta_{ij}=v_{i,r}^{s}+v_{j,s}^r,
\end{equation}
where the symmetry condition, $\omega_{rs}=\omega_{sr}$, is used. The new
parameters, Eq.~(\ref{varthetai}), determine the expected values of partial
degrees, $\langle k_{i,r}^s\rangle$, Eq.~(\ref{kis}).

The two equivalent forms of the Hamiltonian, Eqs.~(\ref{DH1c})
and~(\ref{DH2d}), do not only provide alternative parameterization schemes
for the considered ensemble. They are also helpful in analyzing various
network properties. In particular, Eq.~(\ref{DH2d}) indicates that the
networks may be considered as composed of $K$ independent blocks (i.e. $K$
random graphs with a given node degree sequence, see
Refs.~\cite{PRE2004Park,PRE2006Fronczak}) and the $K$-partite graph
\footnote{$K$-partite graph is a graph where the vertices are partitioned
into $K$ disjoint subsets with the condition that no two vertices in the same
subset are adjacent.}, where $K$ is the number of blocks. Although, the
noticed graph partitioning is quite obvious (especially if one recalls that
the considered ensemble is the ensemble of networks with independent edges),
it is also true that the partitioning is not as straightforward, if one is
only acquainted with Eq.~(\ref{DH1c}). On the other hand, Eq.~(\ref{DH1c}) is
valuable in the sense that by comparing it with the Hamiltonian of the
classical blockmodel, Eq.~(\ref{BH3}), it follows directly that the ensemble
parameters $\theta_{ij}$ provided by Eq.~(\ref{Dthetaij1}) do really define
the degree-corrected blockmodel.

Given the understanding of the networks as composed of $K$ independent
generalized random graphs (blocks) and the $K$-partite graph, the partition
function of the degree-corrected blockmodel can be written in the
multiplicative form:
\begin{equation}\label{DZ1}
Z(\{v_i\!+\!v_j\!+\!\omega_{rs}\})\!=\!\prod_rZ_r\!\prod_{r>s}Z_{rs},
\end{equation}
where $Z_r$ represents the partition function of the block $r$ (see Eq.~(20)
in~Ref.~\cite{PRE2004Park}),
\begin{equation}\label{DZr}
Z_r=\prod_{i<j}\left(1+\delta_{g_ir}\delta_{g_jr}\;e^{(v_i+v_j+\omega_{rr})}\right),
\end{equation}
and $Z_{rs}$ is the partition function of the set of edges between the blocks
$r$ and $s$,
\begin{equation}\label{DZrs}
Z_{rs}=\prod_{i<j}\left(1+\delta_{g_ir}\delta_{g_js}\;e^{(v_i+v_j+\omega_{rs})}\right).
\end{equation}

Now, the average value of the partial degree, $\langle k_{i,r}^s\rangle$, can
be obtained by differentiating the logarithm of the partition function with
respect to $v_{i,r}^s$, i.e.
\begin{equation}\label{Dkis1}
\langle k_{i,r}^s\rangle=\frac{\partial \ln Z}{\partial v_{i,r}^s}=\delta_{g_ir}\sum_{j}\delta_{g_js}\;p_{ij},
\end{equation}
where $p_{ij}$, Eq.~(\ref{Aij1}), is the probability that the vertices $i$
and $j$ are connected,
\begin{equation}\label{Dpij1}
p_{ij}=\frac{\exp\left[v_i+v_j+\omega_{g_ig_j}\right]}{1+\exp\left[v_i+v_j+\omega_{g_ig_j}\right]}.
\end{equation}
Consequently, the average internal and external degrees characterizing nodes
of the considered networks are given by, cf.~Eqs.~(\ref{kiint})
and~(\ref{kiext}),
\begin{equation}\label{Dkint}
\langle k_i^{int}\rangle=\delta_{g_ir}\sum_{j}\delta_{g_jr}\frac{\exp\left[v_{i}+v_{j}+\omega_{rr}\right]}{1+\exp\left[v_{i}+v_{j}+\omega_{rr}\right]},
\end{equation}
and
\begin{equation}\label{Dkext}
\langle k_i^{ext}\rangle=\delta_{g_ir}\sum_{s,s\neq r}\sum_{j}\delta_{g_js}\frac{\exp\left[v_{i}+v_{j}+\omega_{rs}\right]}{1+\exp\left[v_{i}+v_{j}+\omega_{rs}\right]}.
\end{equation}

To proceed further, sparse networks will be considered, in which the
probability of any individual edge is small, $p_{ij}\ll 1$. (The assumption
is reasonable since most of real-world networks are very sparse, i.e. the
number of existing edges is much smaller than the number of edges which could
theoretically exist in the network.) To achieve this, one needs
$e^{v_i+v_j+\omega_{g_ig_j}}\ll 1$ in Eq.~(\ref{Dpij1}), which results in
\begin{equation}\label{Dpij2a}
p_{ij}\simeq\exp[v_i+v_j+\omega_{g_ig_j}]=\exp\left[v_{i,r}^s\right]
\exp\left[v_{j,s}^r\right].
\end{equation}

Eq.~(\ref{Dpij2a}) means that in a sparse degree-corrected blockmodel the
probability of an edge is simply a product of two terms, one for each of the
vertices at either end of the edge. Furthermore, it turns out that these
terms are simply related to the expected partial degrees of the vertices
(similarly to what is found in exponential random graphs with a given node
degree sequence, see~e.g. Eq.~(26) in Ref.~\cite{PRE2006Fronczak}). In order
to see this, one can start with inserting Eq.~(\ref{Dpij2a})
into~(\ref{Dkis1}):
\begin{equation}\label{Dkis2}
\langle k_{i,r}^s\rangle=\exp\left[v_{i,r}^s\right]\sum_j
\exp\left[v_{j,s}^r\right].
\end{equation}
Then, assuming that $s=r$ and summing the last expression over the all
vertices in $r$ one gets
\begin{equation}\label{pom1}
\sum_i\langle k_{i,r}^r\rangle=2\langle E_{rr}\rangle=\left(\sum_i\exp\left[v_{i,r}^r\right]\right)^2,
\end{equation}
from which it follows that
\begin{equation}\label{Dkirr1}
\langle k_{i,r}^r\rangle=\exp\left[v_{i,r}^r\right]\sqrt{2\langle E_{rr}\rangle},
\end{equation}
and
\begin{eqnarray}\label{Dkirs1a}
\langle k_{i,r}^s\rangle\!&\!=\!&\!\delta_{g_ir}\exp\!\left[v_i\!+\!\frac{\omega_{rs}}{2}\right]\!
\sum_{j}\!\delta_{g_js}\exp\!\left[v_j\!+\!\frac{\omega_{rs}}{2}\right]\\\label{Dkirs1b}
\!&\!=\!&\!\exp\!\left[v_{i,r}^r\right]\!\sqrt{2\langle E_{ss}\rangle}\exp\!\left[\!\omega_{rs}\!-\!\frac{(\omega_{rr}\!+\!\omega_{ss})}{2}\!\right]\!.
\end{eqnarray}
Next, by summing Eq.~(\ref{Dkirs1b}) over $i$ one finds for $r\neq s$
\begin{equation}\label{pom2}
\exp\!\left[\frac{(\omega_{rs}\!-\!\omega_{rr})}{2}\!+\!\frac{(\omega_{rs}\!-\!\omega_{ss})}{2}\right]\!=\!\frac{\langle E_{rs}\rangle}{2\sqrt{\langle E_{rr}\rangle\langle E_{ss}\rangle}},
\end{equation}
i.e.
\begin{equation}\label{pom3}
\exp\!\left[\frac{\omega_{rs}\!-\!\omega_{rr}}{2}\right]\!=\!\sqrt{\frac{\langle E_{rs}\rangle}{2\langle E_{rr}\rangle}},
\end{equation}
and the expression for the expected partial degree, Eq.~(\ref{Dkirs1b}), can
be written as
\begin{eqnarray}\label{Dkirs2a}
\langle k_{i,r}^s\rangle&=&\exp\left[v_{i,r}^r\right]\frac{\langle E_{rs}\rangle}{\sqrt{2\langle E_{rr}\rangle}}\\
\label{Dkirs2b}&=&\exp\left[v_{i,r}^s\right]\sqrt{\langle E_{rs}\rangle}.
\end{eqnarray}

\begin{figure}
\includegraphics[width=\columnwidth]{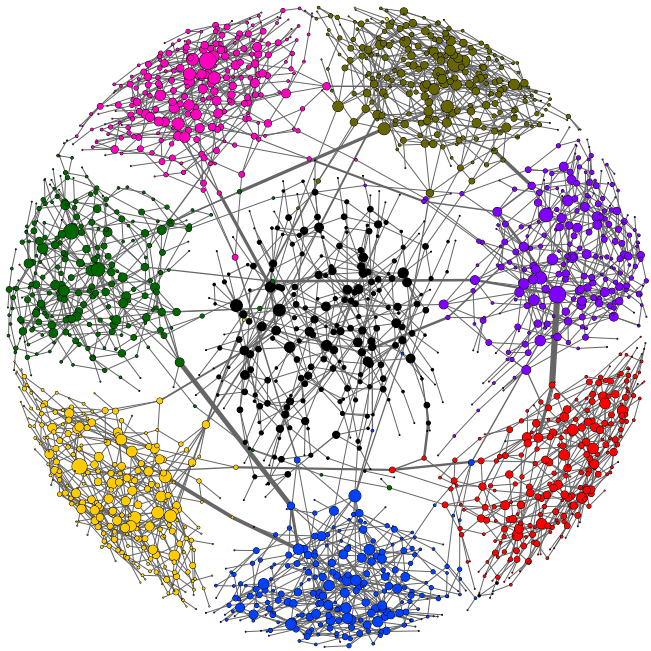}% Here is how to import EPS art
\caption{\label{fig1} (Color online) A Monte Carlo realization of the classical blockmodel with
$N=2048$ nodes divided into $K=8$ groups of equal size $N_r=256$. The color symbols indicate
the group membership, the node size is proportional to the degree, and the edge width is proportional
to the edge betweenness. To get the figure the following values of the ensemble parameters have been used:
$\forall_{r}q_{rr}=3/255$, and $\forall_{r\neq s}q_{rs}=0.006/256$, see Eqs.~(\ref{Bpij}) and~(\ref{Bthetaij}).
The network was visualised using Cytoscape software \cite{cytoscape}.}
\end{figure}

Now, using Eqs.~(\ref{Dkirr1}) and~(\ref{Dkirs2a}) it is straightforward to
obtain the expected internal and external degrees of the nodes,
cf.~Eqs.~(\ref{kiint}) and~(\ref{kiext}),
\begin{eqnarray}\label{Dkiint}
\langle k_i^{int}\rangle &=&\exp\left[v_{i,r}^r\right]\sqrt{2\langle E_{rr}\rangle}\\\label{Dkiext}
\langle k_i^{ext}\rangle &=&\exp\left[v_{i,r}^r\right]\frac{\sum_{s,s\neq r}\langle E_{rs}\rangle}{\sqrt{2\langle E_{rr}\rangle}}.
\end{eqnarray}
Accordingly, it is interesting to note that there is a linear dependence
between the obtained degrees, i.e. for each node $i$ the number of edges
going out from $i$ to nodes in a different block is proportional to the
number of links to nodes in the same block, i.e.
\begin{equation}\label{Dkirs3}
\langle k_{i,r}^s\rangle=\langle k_{i,r}^r\rangle\frac{\langle E_{rs}\rangle}{2\langle E_{rr}\rangle},
\end{equation}
and
\begin{equation}\label{extint}
\langle k_i^{ext}\rangle=\langle k_i^{int}\rangle \frac{\sum_{s,s\neq r}\langle
E_{rs}\rangle}{2\langle E_{rr}\rangle}.
\end{equation}

The scaling relation, Eq.~(\ref{extint}), is an important feature of the
degree-corrected blockmodel, because it relates the model to fractal networks
as described by Song, Havlin and Makse \cite{Nature2005Song,
NaturePhys2006Song}. In Ref.~\cite{Nature2005Song}, the authors argued that
the self-similarity of real-world complex networks with power-law degree
distributions results from the scaling property of the node degrees, which
arises when the networks undergo a renormalization procedure that
coarse-grains their nodes into boxes (blocks). It is remarkable, that the
scaling property reported by Song et al. is similar to the
scale-transformation described by~Eq.~(\ref{extint}). This similarity comes
as a surprise, especially that in this study, contrary to what is suggested
in Ref.~\cite{NaturePhys2006Song}, the relation is not attributed to any
specific network construction procedure. It is an intrinsic feature of the
degree-corrected blockmodel.

Finally, inserting Eqs.~(\ref{Dkirr1}) and~(\ref{Dkirs2a}) into the
expression for the the connection probability, $p_{ij}$, between two vertices
$i$ and $j$, i.e. into Eq.~(\ref{Dpij2a}), one gets that $p_{ij}$ depends on
whether the vertices belong to the same block or not. In the first case, when
$g_i=g_j=r$, one has
\begin{equation}\label{Bpirjr}
p_{ij}=\frac{\langle k_{i,r}^r\rangle\langle k_{j,r}^r\rangle}{2\langle E_{rr}\rangle}.
\end{equation}
In the second case, for $g_i=r$, $g_j=s$, and $r\neq s$, the probability is
given by
\begin{equation}\label{Bpirjs}
p_{ij}=\frac{\langle k_{i,r}^s\rangle\langle k_{j,s}^r\rangle}{\langle E_{rs}\rangle}.
\end{equation}

\section{Numerical simulations}\label{MonteCarloB}

Monte Carlo simulation \cite{2002Snijders,bookNewman} is a numerical method
which is ideally suited to exponential random graphs. In what follows we
briefly describe the method and apply it to the classical blockmodel and its
degree-corrected counterpart.

\subsection{Metropolis-Hastings algorithm for ERGs}

Let us consider the ensemble of exponential random graphs with the
Hamiltonian given by Eq.~(\ref{H0}). Once the parameters $\{\theta_i\}$ in
the Hamiltonian are specified the probability distribution, $P(G)$, which is
given by Eq.~(\ref{PG0}), makes generation of graphs correctly sampled from
the ensemble straightforward using a Metropolis-Hastings type Markov chain
method. In the method, one defines a move-set in the space of graphs and then
repeatedly generates moves from this set, accepting them with probability
\begin{equation}\label{MC1a}
p=1\;\;\;\;\;\mbox{if}\;\;\;\;\;P(G')>P(G),
\end{equation}
where $G'$ is the graph after performance of the move, and
\begin{equation}\label{MC1b}
p=\frac{P(G')}{P(G)}\;\;\;\;\;\mbox{if}\;\;\;\;\;P(G')<P(G),
\end{equation}
while rejecting them with probability $1-p$. Because of the exponential form
of $P(G)$, the acceptance probability which is given by Eq.~(\ref{MC1b}) is
particularly simple to calculate. It can be written as
\begin{equation}\label{MC2a}
p=e^{H(G')-H(G)}=e^{\Delta H}\;\;\;\;\;\mbox{if}\;\;\;\;\;\Delta H<0,
\end{equation}
where
\begin{equation}\label{MC2b}
\Delta H=\sum_r\theta_r\left(m_r(G')-m_r(G)\right).
\end{equation}
Let us also note, that with the help of $\Delta H$, the condition for certain
acceptance of a change, i.e. Eq.~(\ref{MC1a}), becomes
\begin{equation}\label{MC2c}
p=1\;\;\;\;\;\mbox{if}\;\;\;\;\;\Delta H>0.
\end{equation}

The choice of the right move-set depends on the set of all possible network
realizations, $\{G\}$, underlying the studied ensemble. The example move-sets
are: i) addition and removal of edges between randomly chosen vertex pairs
for the case of graphs which do not have a fixed number of edges; ii)
movement of edges randomly from one place to another for the case of fixed
edge numbers but variable degree sequence; iii) edges swaps of the form
$\{(v_1,w_1),(v_2,w_2)\}\rightarrow \{(v_1,w_2),(v_2,w_1)\}$ for the case of
fixed degree sequence, where $(v_1,w_1)$ denote an  edge from vertex $v_1$ to
vertex $w_1$. Monte Carlo numerical simulations of this type are simple to
implement and appear to converge quickly allowing one to study quite large
graphs.

\begin{figure}
\includegraphics[width=\columnwidth]{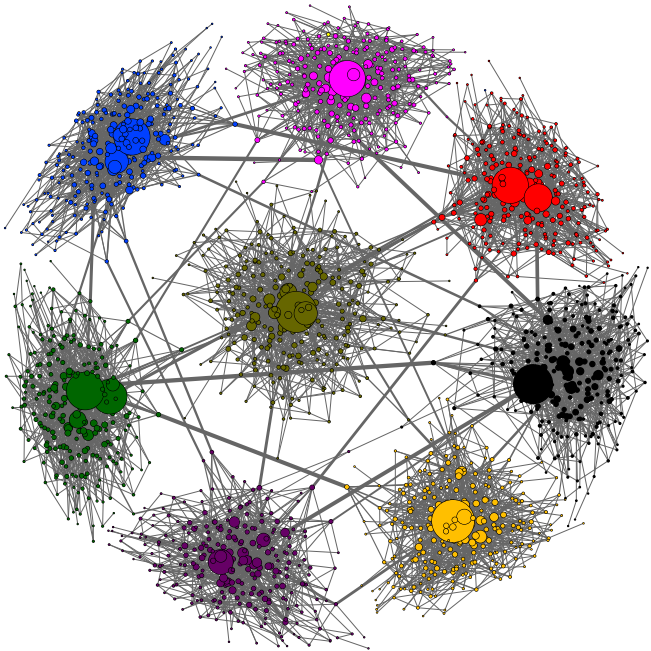}% Here is how to import EPS art
\caption{\label{fig2} (Color online) A Monte Carlo realization of the degree-corrected
blockmodel with scale-free communities. The network consists of $N=2048$ nodes divided into
$K=8$ groups of equal size. The color indicates the partition of the nodes into groups,
and the size of the nodes is proportional to the degree. The edge width is proportional to the edge
betweenness. To perform numerical simulations the following values of the ensemble parameters have been used:
i. The expected internal degrees have been independently drawn from the power law distribution with the characteristic exponent $\gamma=3$,
ii. The minimal value of the expected degree was assumed to be $2$, iii. The expected numbers of interblock
connections were taken to be $\forall_{r\neq s} \langle E_{rs}\rangle=1$.
The network was visualised using Cytoscape software \cite{cytoscape}.}
\end{figure}

\subsection{Girvan-Newman benchmark}

To be concrete, let us discuss the Metropolis algorithm for the classical
blockmodel, which corresponds to the famous Girvan-Newman benchmark for
testing community detection methods. At the beginning one assumes that the
number of groups, $K$, and the number of vertices in each group $N_r$ are
known, and $\sum_{r=1}^KN_r=N$. Groups to which the nodes belong are also
specified, likewise the connection probabilities, $q_{rs}$ between and within
the blocks.

Having the ensemble parameters, the ensemble construction procedure proceeds
through the following steps:
\begin{itemize}
\item[1.] At the beginning one creates any simple graph (i.e. its
    adjacency matrix) with a given number of nodes, $N$. The starting
    configuration may be, for instance, the edgeless graph.
\item[2.] Next, in the following time steps, one randomly chooses a
    matrix element, $A_{ij}(G)$, to be considered for change. For the
    case, when $A_{ij}(G)=1$ ($0$, respectively) one considers deletion
    (addition) of the edge, i.e. $A_{ij}(G')=0$ ($1$, respectively). This
    corresponds to the move-set: addition and removal of edges between
    randomly chosen vertex pairs. Whether the change is accepted depends
    on $\Delta H$, see Eq.~(\ref{MC2b}). In the classical blockmodel,
    since $H(G)=\sum_{i<j}\omega_{g_ig_j}A_{ij}(G)$, cf.~Eq.~(\ref{BH1}),
    one has
    \begin{equation}\label{dHB}
    \Delta H=\pm\omega_{g_ig_j},
    \end{equation}
    with the upper (lower) sign relating to addition (deletion,
    respectively) of an edge. Therefore, the acceptance criteria,
    Eqs.~(\ref{MC2a}) and (\ref{MC2c}), depend on the sign of the
    ensemble parameters $\omega_{rs}$, which are given by
    Eq.~(\ref{Bthetaij}).
\item[3.] The updating of elements $A_{ij}$ should be continued until the
    network observables stabilise around their mean values. In the case
    of the classical blockmodel, the average numbers of edges, $\langle
    E_{rr}\rangle$ and $\langle E_{rs}\rangle$, should place itself
    around the values, which are given by Eqs.~(\ref{BErr}) and
    (\ref{BErs}). Once numerical simulations stabilise, graphs which
    appear in the course of subsequent updates of the adjacency matrix
    appear to be correctly sampled network realizations of the studied
    ensemble.
\end{itemize}

A realization of the classical blockmodel with $N=2048$ nodes divided into
$K=8$ groups of equal size is shown in Fig.~\ref{fig1}.

\subsection{Degree-corrected blockmodel}

The construction procedure of the degree-corrected blockmodel follows the
same steps as in the classical blockmodel described in the previous
subsection. The only difference is that the change of the Hamiltonian
occurring during the addition or removal of an edge between $i$ and $j$ is,
cf.~Eq.~(\ref{Dthetaij2}),
\begin{equation}\label{dHDC}
\Delta H=\pm(v_{i,r}^s+v_{j,s}^r).
\end{equation}
Therefore, to perform Monte Carlo simulations of the degree corrected
blockmodel, first, one has to determine the set of parameters $\{v_{i,r}^s\}$
underlying the network ensemble with the desired (e.g. scale-free) degree
distribution.

Thus, let us start with specifying the parameters $v_{i,r}^r$ which
characterize the expected internal degrees of the nodes, $\langle
k_i^{int}\rangle=\langle k_{i,r}^r\rangle$. In order to do it we have to
assume the desired node degree sequence within each block, $\{\langle
k_{1,r}^r\rangle, \langle k_{2,r}^r\rangle,\dots,\langle
k_{N_r,r}^r\rangle\}$, where the numbers $\langle k_{i,r}^r\rangle$ can be
drawn from an arbitrary degree distribution. Then, the parameters,
$\{v_{i,r}^r\}$, can be calculated from Eq.~(\ref{Dkiint}):
\begin{equation}\label{virr}
v_{i,r}^r=\ln\left[\frac{\langle k_{i,r}^r\rangle}{\sqrt{\sum_{i=1}^{N_r}\langle k_{i,r}^r\rangle}}\right].
\end{equation}

To determine the parameters $\{v_{i,r}^s\}$ for $r\neq s$, the average
numbers of interblock connections, $\langle E_{rs}\rangle$, has to be given.
Then, once $\forall_{r\neq s}\langle E_{rs}\rangle$ are known, the
parameters, $\{v_{i,r}^s\}$, can be obtained from Eqs.~(\ref{Dkirs2b}),
(\ref{Dkiint}), and (\ref{Dkirs3}):
\begin{eqnarray}\label{virs}
v_{i,r}^s&=&\ln\left[\frac{\langle k_{i,r}^s\rangle}{\sqrt{\langle E_{rs}\rangle}}\right]=
\ln\left[\frac{\langle k_{i,r}^r\rangle\sqrt{\langle E_{rs}\rangle}}{\sum_{i=1}^{N_r}\langle k_{i,r}^r\rangle}\right]\\
&=&v_{i,r}^r+\ln[\sqrt{\langle E_{rs}\rangle}].
\end{eqnarray}

In Fig.~\ref{fig2} a realization of the degree-corrected blockmodel is shown.
The network has $N=2048$ nodes which are divided into $K=8$ communities of
equal size. The node degree distribution within the communities is
scale-free, $\forall_r P(\langle k_{i,r}^{r}\rangle) \sim \langle
k_{i,r}^{r}\rangle^{-\gamma}$, with the characteristic exponent $\gamma=3$.

\section{Summary}\label{Conclude}

In this paper, we have described an exponential random graph approach to
networks with community structure. We mainly built upon the idea of the
blockmodel, although other ideas regarding network structure have been also
exploited. Two kinds of the network structural Hamiltonians have been
considered: the first one corresponding to the classical blockmodel, and the
second one corresponding to its degree-corrected version. In both cases, a
number of analytical predictions about various network properties was given.
In particular, it was shown that in the degree-corrected blockmodel, node
degrees display an interesting scaling property, that is similar to the
scaling feature of the node degrees in fractal (self-similar) real-world
networks. A short training in Monte Carlo simulations of the models was also
given in the hope of being useful to others working in the field of networks
with community structure.

\section{Acknowledgments}

This work was supported by the Foundation for Polish Science under grant
no.~POMOST/2012-5/5.

%\bibliography{blockmodels}% Produces the bibliography via BibTeX.

\end{document}